\documentclass[usenatbib,usedcolum,usegraphicx]{mn2e}

\title[
Variability-selected QSO candidates in OGLE-II Galactic Bulge fields
      ]{
Variability-selected QSO candidates in OGLE-II Galactic Bulge fields
 }

\author[             T. ~Sumi et al.
       ]
       {             T. ~Sumi,$^1$ P. R.~Wo\'{z}niak,$^2$ L.~Eyer,$^3$  
        A.~Dobrzycki$^4$  A.~Udalski$^5$, M. K.~Szyma{\'n}ski$^5$,    \newauthor
        M.~Kubiak$^5$, G.~Pietrzy\'nski$^{5,6}$, 
        I.~Soszy\'nski$^5$, K.~\.Zebru\'n$^5$,  
        O.~Szewczyk$^5$   \newauthor \& \L.~Wyrzykowski$^5$ \\
    $^1$Princeton University Observatory, Princeton, NJ 08544-1001, USA,
    e-mail: sumi@astro.princeton.edu \\ 
    $^2$Los Alamos National Laboratory, MS-D436, Los Alamos, NM 87545,
    e-mail: wozniak@lanl.gov\\
    $^3$Observatoire de Gen\`eve, CH 1290 Sauverny, Switzerland, e-mail: laurent.eyer@obs.unige.ch\\
    $^4$European Southern Observatory, D-85748 Garching bei M\"{u}nchen, Germany, e-mail: adam.dobrzycki@eso.org\\
    $^5$Warsaw University Observatory, Al.~Ujazdowskie~4, 00-478~Warszawa, Poland;\\
~~~e-mail: (udalski,msz,mk,pietrzyn,soszynsk,zebrun,szewczyk,wyrzykow)@astrouw.edu.pl\\
$^6$ Universidad de Concepci{\'o}n, Departamento de Fisica, Casilla 160--C, Concepci{\'o}n, Chile\\
}
\date{Accepted 
      Received
      in original form}

\pagerange{\pageref{firstpage}--\pageref{lastpage}}

\begin{document}
\maketitle
\label{firstpage}

\begin{abstract}

We present 97 QSO candidates in 48 Galactic Bulge fields
of the Optical Gravitational Lensing Experiment II (OGLE-II)
covering $\sim11$ square degrees, which are selected via their
variability. 
We extend light curves of variable objects which were
detected in 3-year baseline in the OGLE-II variable star catalog
to 4th year.
We search for objects which are faint ($16<I_0<18.5$) and
slowly variable during 4 years in this catalog by
using the variogram/structure function. 
Finding the QSOs in such
stellar-crowded and high extinction fields is challenging, but should
be useful for the astrometric reference frame.  Spectroscopic
follow-up observations are required to confirm these QSO candidates.
Follow-up observations are being prepared for four of these fields
(BUL\_SC1, 2, 32 and 45).  Follow-up observations for other fields are
strongly encouraged. The complete list and light curves of all 97 
candidates are available in electronic format at 
http://www.astro.princeton.edu/~sumi/QSO-OGLEII/.

\end{abstract}


\section{Introduction}

QSOs have been found by a variety of different methods, mostly using
optical photometry and spectroscopy, but also through their radio
morphology and X-ray emission properties.

QSO searches were usually avoided in areas where source crowding was
a problem, for example the Galactic plane or the direction of nearby
galaxies, such as the Magellanic Clouds. High spatial resolution of the
Chandra X-ray Observatory limited source confusion problems for the
X-ray based searches. \cite{dob02} found
several quasars behind the LMC and SMC, including four quasars behind
the dense parts of the LMC bar by spectroscopic observations of
the OGLE optical counterpart of Chandra X-ray sources.
 
Another efficient method of searching for quasars is based on optical
variability
(\citealt{ha83}; \citealt{ver95}; or more recently \citealt{bru02};
\citealt{ren04}).  Several gravitational microlensing survey groups
have been monitoring hundreds of millions of stars toward dense
regions such as the Galactic centre, the Galactic disc and/or the
Magellanic clouds (EROS: \citealt{der99}; OGLE: \citealt{uda03};
MACHO: \citealt{alc00}; MOA: \citealt{bon01}).  
\cite{eye02} searched QSO candidates towards the
Magellanic clouds via variability by using the second phase of the
OGLE experiment
\footnote{see {\tt http://www.astrouw.edu.pl/\~{}ogle} or {\tt
http://bulge.princeton.edu/\~{}ogle}} (OGLE-II; \citealt{uda97})
variable star catalog (\citealt{zeb01}). \cite{dob03} did follow-up
spectroscopic observations of 12 of these quasar candidates in the
Small Magellanic Cloud fields and confirmed 5 as QSOs.
They also found the color information is useful to select QSOs. 
\cite{geh03} found about 47 QSOs behind the Magellanic Clouds, by
selecting the candidates via variability in the MACHO database and
follow-up spectroscopic observations.

In this work, following \cite{eye02}, we search for QSO candidates via
variability in OGLE-II Galactic Bulge fields (\citealt{uda02}).  Up to
now, no QSOs have been found in these regions. The main motivation for
this study is that quasars may provide background sources which fix
the astrometric reference frame.  Kinematic studies of stars
(\citealt{sum03,sume04}) would greatly benefit from having absolute
astrometry.  These QSOs will also be useful for the future astrometric
program, such as  Space Interferometry Mission (SIM) and GAIA.  
However, finding QSOs in these fields is more challenging
than in Magellanic Cloud fields because of the higher stellar density
and higher extinction.

%
%
%

In \S\,\ref{sec:data} we describe the data.  We select QSO candidates
in \S\,\ref{sec:select}. Discussion and conclusion are given in
\S\,\ref{sec:disc}.

\section{DATA}
\label{sec:data}

We use the $VI$ photometric maps of OGLE-II fields
(\citealt{uda02}), which contain $VI$ photometry and astrometry of
$\sim 30$ million stars in the 49 Galactic Bulge fields. Positions of
these fields (BUL\_SC1 $\sim$ 49) can be found in \cite{uda02}. We do
not use BUL\_SC44 in this work because its extinction is too high
($A_V>4$). The photometry is the mean photometry from a few hundred 
measurements in the $I$-band and several measurements in $V$-band collected 
between 1997 and 2000.  Accuracy of the zero points of the photometry is about
0.04 mag. A single 2048 $\times$ 8192 pixel frame covers an area of
0.24 $\times$ 0.95 $deg.^2$ with a pixel size of 0.417 arcsec.
Details of the instrumentation setup can be found in \cite{uda97}.

We also use the light curves of objects in the OGLE-II Galactic Bulge
variable star catalogue (\citealt{woz02}) which contain 213,965
variable objects.  The variable object detection and the photometry
have been done by using the Difference Image analysis (DIA)
(\citealt{ala98}; \citealt{ala00}; \citealt{woz00}) for the data
collected during the first 3 years of the OGLE-II project.  We
extended the light curves to the fourth year (2000) using the method
of \cite{woz02}; however new variable object detection is not
attempted for the additional data.

\section{QSO candidates selection}
\label{sec:select}

We cross-referenced 213,965 variable objects in the variable star
catalogue (\citealt{woz02}) with objects in OGLE-II photometric maps
(\citealt{uda02}), which are well calibrated, and extracted their $V$
and $I$ magnitudes.  205,473 of them have counterparts in the
photometric maps.  We use these magnitudes in the following
analysis.

\begin{figure}
\begin{center}
\includegraphics[angle=0,scale=0.5,keepaspectratio]{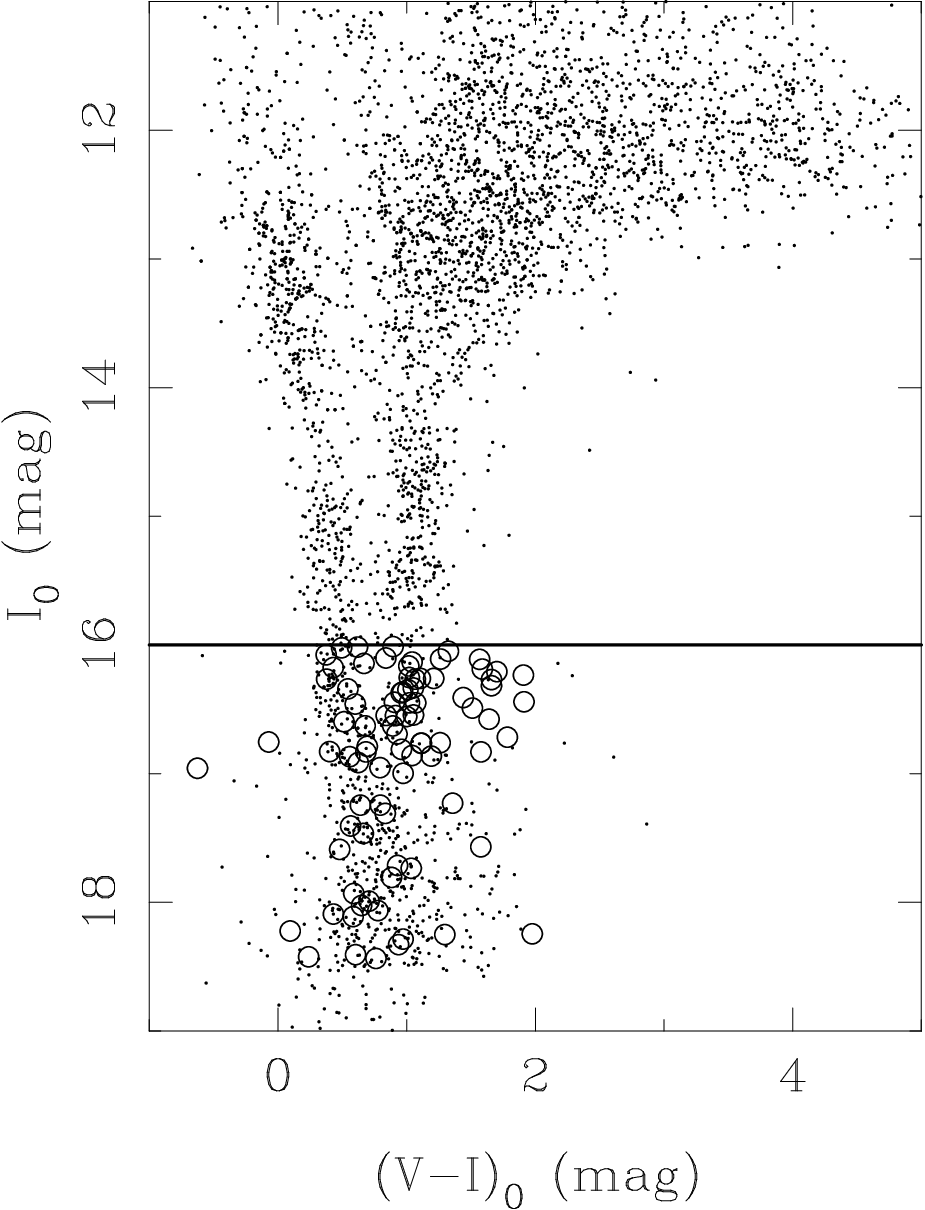}
\caption{
Extinction corrected Colour Magnitude Diagram of variable stars 
in BUL\_SC1 (dots) and QSO candidates in all fields (open circles).
 Candidates are searched for $I_0 <16$.
  \label{fig:cmd}
  }
\end{center}
\end{figure}

We chose 30,219 objects which are fainter than $I_0=16$ mag, where
$I_0$ is the $I$-band magnitude after extinctions are corrected by
using the extinction map in these fields (\citealt{sum04}).  We show
the Colour Magnitude Diagram (CMD) of one of these fields in
Fig.~\ref{fig:cmd}. The detection limit in the variable star catalogue
is about $I=19.5$ mag, which corresponds to extinction corrected
magnitudes of $17 \sim 18.5<I_0$, depending on the level of extinction
in each field.  
The number density of QSO  brighter than $i=19.1$ mag is $9.6/deg.^2$
(\citealt{van04}), where the Sloan $i$ is almost similar to the OGLE $I$.
So in this magnitude range we expect $20 \sim 30$ QSOs
in our 48 fields covering $\sim 11$ square degrees taking
the extinction into account.

There are many artifacts in the variable star catalogue. The light
curves of objects around bright stars are affected by the long wings
of the Point Spread Function (PSF) of that bright star, and then they
resemble that star's variability. Even around a star with constant 
brightness, the wings can affect nearby faint stars due to variations 
of seeing and the PSF. This effect causes the systematic gap in the brightness
of faint stars around a bright star between in 1998 and in 1999 due to the 
re-aluminization of the mirror, which changed the size and/or shape 
of the PSF.
Such light curves with a systematic gap are one of the major
contaminations in our QSO selection. So we masked the stars, around
the bright stars including saturated stars.  Saturated stars are not
in the star catalogue.  To deal with them we counted the number of
continuous pixels, $N_{\rm sat}$ whose ADU value is larger than 12,000
on the OGLE-II template image.  Then we masked the area centered at
these pixels with a radius of $rad= 4 + 5 \times \sqrt{N_{\rm sat}}$.
This removed $62 \%$ of variable stars though only $11\%$ of the area
in total is masked.  At this point, 11,535 light curves still
remain.

\begin{figure}
\begin{center}
\includegraphics[angle=-90,scale=0.33,keepaspectratio]{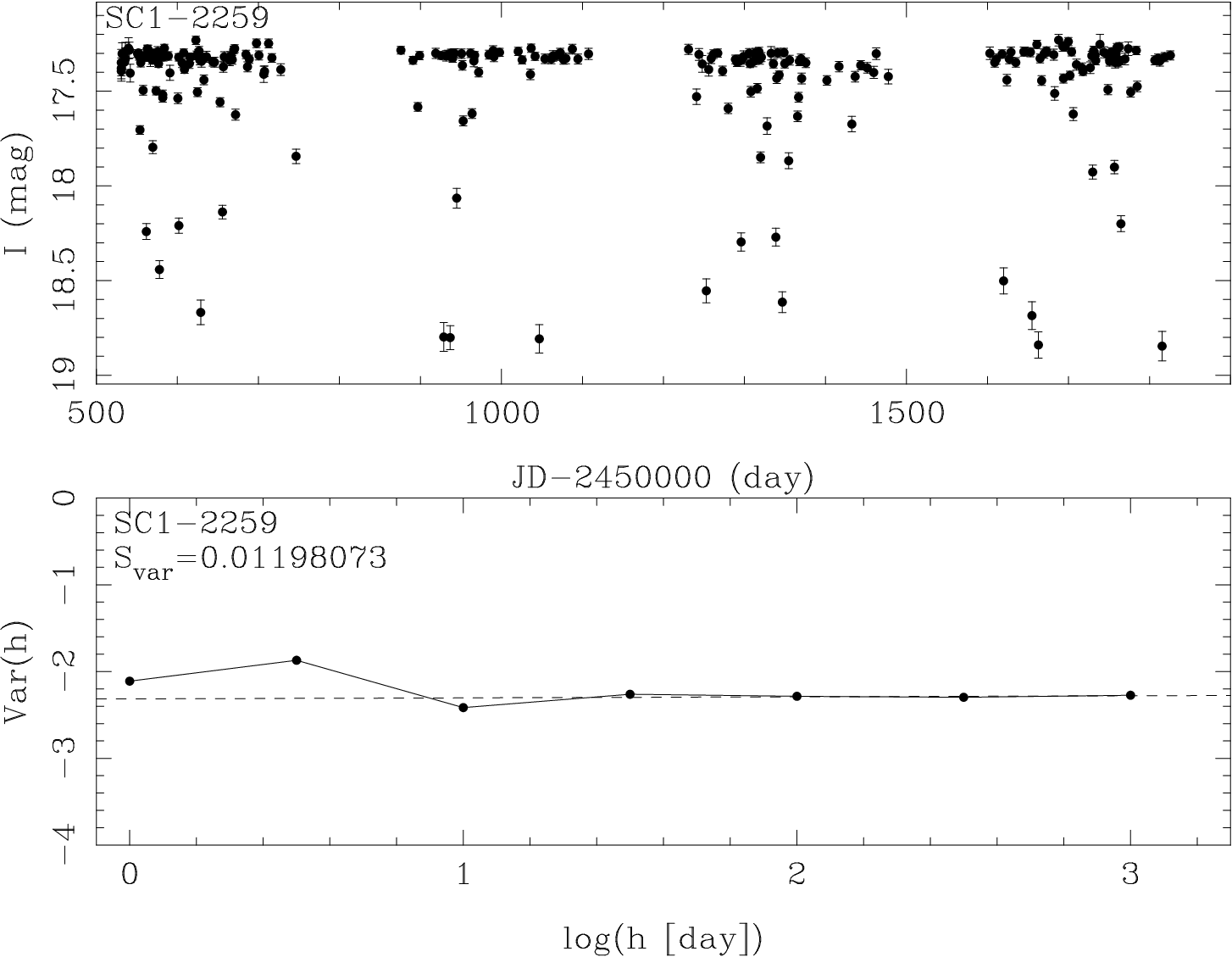}
\includegraphics[angle=-90,scale=0.33,keepaspectratio]{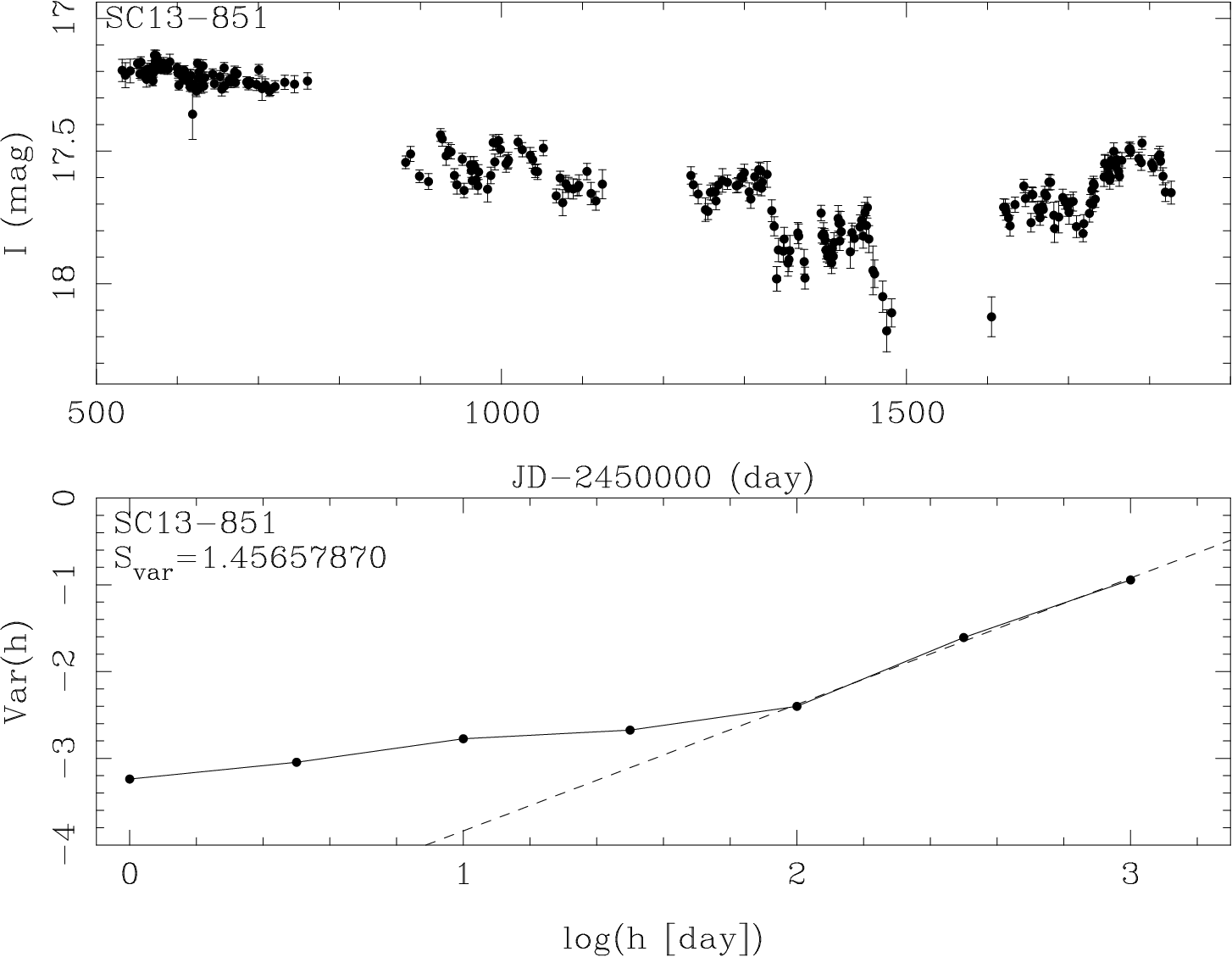}
\caption{
  \label{fig:Var}
Sample light curves (first and third panel) and variogram/structure function 
$Var(h)$ as a function of log$h$ (second and 4th panel, dot with solid line)
for objects which passed the Mask around bright stars.
The top two and bottom two panels are objects which failed (sc1-2259)
and passed (sc13-851) the criteria, $S_{\rm var}\ge0.09$ and AoV algorithm $<50$.
In the second and 4th panels, dashed lines represent the best fit for the three points
(log$h=2, 2.5, 3$) whose slopes $S_{\rm var}$ are written in the panels.
  }
\end{center}
\end{figure}

Following \cite{eye02}, we use the variogram/structure function
(\citealt{eye99}, \citealt{hug92}) to find objects with increasing
variability on longer time-scales.  We measure the slope of the
variogram as follows.  For all possible pairs of measurements,
differences of times, $h_{i,j} = t_j-t_i$ and square of difference of
magnitudes, $(I_j-I_i)^2$ are computed. For a given bin of log$h$, the
median of $(I_j-I_i)^2$ is estimated, and is noted as $Var(h)$. This
gives a spread of the variation of $I$ in this time-scale.  We bin
log$h$ for three parts divided by 1.5, 2, 2.5, 3, then measure the
slope, $S_{\rm var}=dVar(h)/d{\rm log}h$.  The larger $S_{\rm var}$, the
larger the variability on larger time-scales.  We rejected light
curves with $S_{\rm var}<0.09$, which is a slightly more conservative
value than the 0.1 used in \cite{eye02}. We search the periodicity in
the light curves by using the AoV algorithm (\citealt{schw89}).
Then we reject the light curves in which strong periodicity is detected 
with the AoV periodogram of larger than 50. This value was chosen
empirically.  We show two sample light curves and $Var(h)$ as a 
function of log$h$ in Fig. \ref{fig:Var}.
After this procedure, 3,201 light curves remain.

The remaining light curves, were inspected by eye. First we rejected
apparent periodic and semi-periodic variables which couldn't
be rejected by the previous procedure.  After this 500 light curves
are left.

The main contamination after this selection are high proper motion
objects and artifacts around bright Variable stars. We rejected them
as follows.

When an object moves from the position at the template image, DIA
produces a pair of positive and negative residuals on the subtracted
image. The flux of these residuals increases as a function of the
displacement of positions of the objects (\citealt{eye01};
\citealt{sos02}).  We cross-referenced with the proper motion catalog
(\citealt{sume04}) to check if they have high proper motion or if there
is a high proper motion object nearby.  Even if its proper
motion is not so high, a bright nearby star can produces a pair of significant 
residuals and constantly increase or decrease the brightness of the 
nearby faint stars.  To reject them we looked
at the light curves of all neighboring stars in OGLE-II photometric maps
within 20 pixels around the object, which were made by using DIA.
Then we checked whether they have a companion with an opposite 
(increasing or decreasing brightness) light curve around them.

In this process, we also rejected the light curves which have
neighbours with similar light curves. Usually they are affected by
nearby bright variable stars but were not masked and not clearly 
identified in previous procedures because the shape of their light curves are
distorted from the original variable due to the combination of the
other artifacts or low S/N.
The combination of the effect of the variable star and the high proper
motion object resemble the QSO light curve.

\begin{figure*}
\begin{center}
\includegraphics[angle=0,scale=0.85,keepaspectratio]{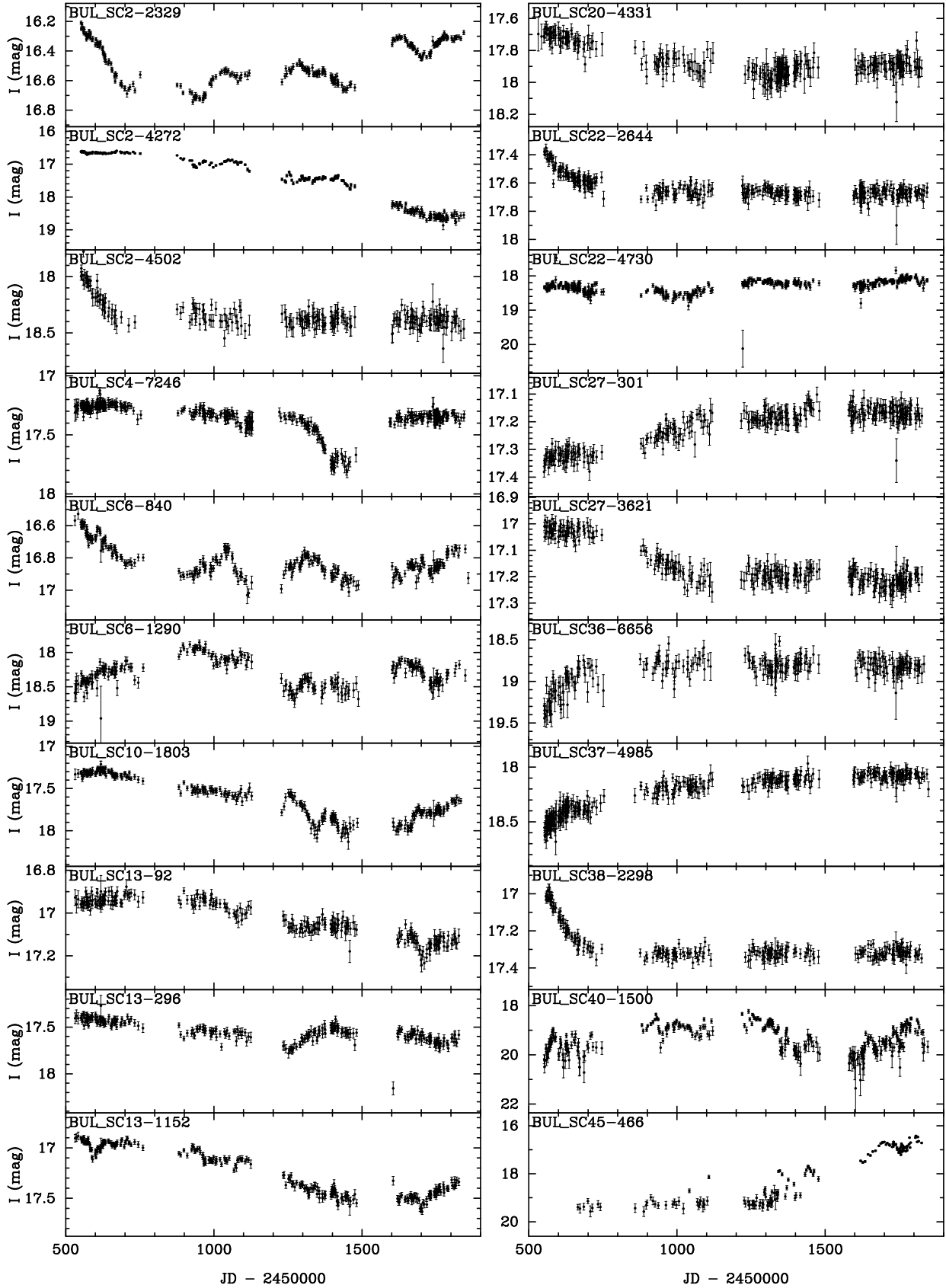}
\caption{
20 sample $I$-band light curves in 97 QSO candidates. 
The light curves of all 97 candidates are available
in electronic format at http://www.astro.princeton.edu/~sumi/QSO-OGLEII/.
  \label{fig:lc} }
\end{center}
\end{figure*}

Finally we have 97 QSO candidates. 
We summarize our selection criteria and the number of remaining 
candidates after each selection criterion in Table \ref{tbl:Select}.
We show 20 sample light curves in
Fig.~\ref{fig:lc} and list them in Table~\ref{tbl:cand}, where the
objects with $V-I = (V-I)_0= 9.999$ don't have $V$-band photometry.
The complete list and light curves of all 97 candidates are available
in electronic format at http://www.astro.princeton.edu/~sumi/QSO-OGLEII/. 
We also show their position on the CMD as open circles in
Fig.~\ref{fig:cmd}.

As we can see in Fig.~\ref{fig:cmd}, many of our candidates are
relatively bright ($I_0=16\sim 17$), which is not expected from a
genuine quasar Luminosity Function. This means the contamination by
irregular variable stars is higher in this range than expected. In
this work we didn't put any cut by their colour, because of the
following reasons.  For blue objects, there are only few candidates
with $(V-I)_0 <0.4$ (\cite{dob03}). The Be stars which were a major
contamination in the previous work (\citealt{eye02}) are above the
threshold brightness $I_0<16$ mag.  For the redder objects, the
extinction may be underestimated for background objects because the
extinctions are estimated only up to the Galactic centre in
\cite{sum04}. Some objects don't have $V$ photometry.

\begin{table}
 \caption{The Number of remaining candidates after each selection criterion. 
 \label{tbl:Select}}
    \begin{tabular}{lr}\\
    Criteria & Candidates  \\
  \hline
All Variable objects                              & 213,965 \\
Faint (Extinction corrected I, $I_0>16$ mag)      & 30,219 \\
Mask around bright stars                          & 11,535 \\
$S_{\rm var}\ge0.09$ and  AoV periodogram $<50$   & 3,201 \\
Eye inspection 1, cut apparent periodic variables & 500 \\
Eye inspection 2, cut high proper motion objects  & 97 \\
\end{tabular}\\
\end{table}

\begin{table*}
 \caption{The list of the first 10 sample of 97 QSO candidates. 
 The complete list of all 97 candidates is available in electronic 
format at http://www.astro.princeton.edu/~sumi/QSO-OGLEII/.
Columns are field number, identification number as in Wo\'{z}niak et
al. (2001), ID$_{\rm v}$, and as in Udalski et al. (2002), ID,
Positions on CCD in pixel, $x$ and $y$, Equatorial coordinates
R.A. and Dec. in equinox 2000, $I$-band magnitude and colour, $I$,
$V-I$ and extinction corrected magnitude and colour, $I_0$, $(V-I)_0$,
Variogram slope, $S_{\rm var}$.  The objects with $V-I = (V-I)_0= 9.999$
don't have $V$-band photometry. 
 \label{tbl:cand}}
    \begin{tabular}{lrrrrccccccc}\\
    field & ID$_{\rm v}$ &  ID 
  & x (pixel) & y (pixel) & R.A. (2000) & Dec. (2000)
  & $I$ & $V-I$ & $I_0$ & $(V-I)_0$  & $S_{\rm var}$
    \\
  \hline
BUL\_SC1 & 2411 & 468485 & 1350.47 & 4479.39 & 18:02:42.91 & -29:55:02.7 & 18.828 &  1.195 & 18.091 &  0.431 & 0.19\\
BUL\_SC2 & 2329 & 304696 &  937.12 & 3632.55 & 18:04:25.85 & -28:55:46.9 & 17.118 &  2.614 & 16.441 &  1.912 & 0.94\\
BUL\_SC2 & 2423 & 506014 & 1282.18 & 3824.72 & 18:04:36.73 & -28:54:27.5 & 17.416 &  1.789 & 16.764 &  1.113 & 0.68\\
BUL\_SC2 & 3191 & 337864 &  947.12 & 4930.96 & 18:04:26.21 & -28:46:49.6 & 19.155 &  1.379 & 18.407 &  0.603 & 0.15\\
BUL\_SC2 & 3340 & 136127 &  369.57 & 5145.97 & 18:04:08.05 & -28:45:20.2 & 17.081 &  1.691 & 16.373 &  0.957 & 0.13\\
BUL\_SC2 & 4136 & 165056 &  138.64 & 6347.10 & 18:04:00.87 & -28:37:03.0 & 17.087 &  1.317 & 16.341 &  0.544 & 0.43\\
BUL\_SC2 & 4272 & 164645 &  385.54 & 6631.55 & 18:04:08.64 & -28:35:05.5 & 17.289 &  1.941 & 16.449 &  1.070 & 2.31\\
BUL\_SC2 & 4502 & 775796 & 1942.63 & 6850.26 & 18:04:57.55 & -28:33:35.6 & 18.521 &  1.625 & 17.806 &  0.883 & 0.25\\
BUL\_SC4 & 3598 &  84694 &  129.22 & 3250.07 & 17:54:07.22 & -29:49:30.2 & 18.059 &  1.793 & 16.868 &  0.558 & 0.27\\
BUL\_SC4 & 6483 & 717606 & 1960.77 & 5624.77 & 17:55:05.47 & -29:33:08.2 & 18.841 &  9.999 & 17.697 &  9.999 & 0.09\\
\end{tabular}\\
\end{table*}

\newpage

\section{Discussion and Conclusion}

\label{sec:disc}

We have selected 97 QSO candidates via variability from OGLE-II DIA
variable star catalogue in the Galactic Bulge fields.

Finding the QSOs in such crowded and high extinction fields is
challenging, but should be useful for the astrometric zero-point in
these fields and for the study of the foreground dust. It is
especially useful for the study of the bar structure by using the
stellar proper motions in these fields (\citealt{sume04}).
These QSOs will also be useful for the future astrometric program,
such as  Space Interferometry Mission (SIM) and GAIA.

Spectroscopic follow-up observations are required to confirm these
QSOs. The multi object and wide field spectrographs, such as VIMOS/VLT
and LDSS/Magellan, are suitable for this purpose.  We are planning to
carry out spectroscopic follow-up observations in 4 of these fields
(BUL\_SC1, 2, 32 and 45) by using VIMOS/VLT.  Follow-up observations
in other fields are strongly encouraged.

As we can see in Fig.~\ref{fig:cmd}, there are more candidates at
relatively bright ($I_0=16\sim 17$) range, which is not expected from a
genuine quasar Luminosity Function. This means the contamination by
irregular variable stars is higher in this range than the fainter 
range. So, the efficiency would be higher in these fainter  ($I_0>17$)
candidates.
There are several objects which may be microlensing events, i.e. 
SC2-3191, SC2-4502, SC14-4009, SC19-4350, SC24-3042, SC30-3977, SC38-2298,
and SC39-5030. We retained these objects because we do not know 
they are true microlensing or not.  Also there are several
objects that have light curves too similar to each other to be QSOs,
i.e. SC4-7246, SC13-332, SC13-775, SC20-3616, SC32-1795
SC42-3375, SC42-3772, SC46-742. They have relatively constant baseline
with W or V-shape dent in light curves. They could be new a type of variable
star. We did not reject them because we do not know what they are.

The current dataset is not optimal to find QSOs. The OGLE-II
Galactic Bulge data are relatively shallower than the OGLE-II
Magellanic Cloud data, which are optimized to find microlensing
events.  Variable objects were only searched during the first 3
years for some reason. We can improve the sensitivity by including the
4th year of OGLE-II and OGLE-III to increase the time-scale and
the survey area.

The main contamination in this work was artifacts around bright stars
or from high proper motion stars rather than Be stars as in the previous
work (\citealt{eye02}).  Because the foreground stars are closer than
Magellanic Clouds, most Be stars are above the threshold
brightness.  On the other hand, due to much higher stellar density,
the artifacts are increased.

The work to avoid such artifacts in the photometry on the subtracted
images is underway. If such a new photometry were to be done, then the
efficiency in finding QSOs would be increased.


\section*{Acknowledgments}

We are grateful to Prof.~B.~Paczy\'{n}ski for helpful comments and
discussions. We thank Prof.~P.~Wiita and Prof.~Z.~Ivezic  for 
informative discussions and comments on this article. 
We thank J. Tan for carefully reading the manuscript.
T.S. acknowledges the financial support from
the JSPS.  L.E. is indebted to Prof.~B.~Paczy\'nski for a scientific
stay in Princeton in 2004.  This work was partly supported with the
following grants to Prof.~B.~Paczy\'nski: NSF grant AST-0204908, and
NASA grant NAG5-12212.  OGLE project is partly supported by the Polish
KBN grant 2P03D02124 to A. Udalski.

\label{lastpage}
\clearpage

\end{document}